\documentclass{article}
\usepackage{graphicx}
\usepackage[centertags]{amsmath}
\usepackage{amsfonts}
\usepackage{amssymb}
\usepackage{amsthm}
\newcommand{\rd}{\mathrm{d}}
\newcommand{\ri}{\mathrm{i}}

\title{Mechanics of the Infinitesimal Gyroscopes on the Mylar Balloons
and Their Action-Angle Analysis}

\author{Vasyl~Kovalchuk$^{1}$, Iva{\"\i}lo~M.~Mladenov$^{2}$\\
$^{1}$ Institute of Fundamental Technological Research\\
Polish Academy of Sciences\\
$5^{\rm B}$, Pawi\'{n}skiego str., 02-106 Warsaw, Poland\\
$^{2}$ Institute of Biophysics and Biomedical Engineering\\
Bulgarian Academy of Sciences\\
Acad.~G.~Bonchev Str., Bl.~21, 1113 Sofia, Bulgaria\\
e-mail: vkoval@ippt.pan.pl,\quad mladenov@bio21.bas.bg}

\begin{document}

\maketitle

\begin{abstract}
Here  we apply the general scheme for description of the mechanics of infinitesimal bodies in the Riemannian spaces to the examples of geodetic and non-geodetic (for two different  model potentials) motions of infinitesimal rotators on the Mylar balloons. The structure of partial degeneracy is investigated with the help of the corresponding Hamilton-Jacobi equation and action-angle analysis. In all  situations it was found that for any of the six disjoint regions in the phase space among the three action variables only two of them are essential for the description of our models at the level of the old quantum theory (according to the Bohr-Sommerfeld postulates). Moreover, in both non-geodetic models the action variables were intertwined with the quantum number $N$ corresponding to the quantization of the radii $r$ of the inflated Mylar balloons.
\end{abstract}

\section*{Introduction}

\noindent The general formulation of mechanics of extended metrically- or affinely-rigid bodies in Euclidean spaces was studied in details in some of our previous papers (see, e.g., \cite{all_14,all_10,all_17,all_11,all_12}). The situation when Euclidean/affine space is replaced by a differential manifold equipped with geometry given by the metric tensor, affine connection, or both of them (interrelated or not) was mainly covered in \cite{jjs_76,all_06}. In the present paper we are following the general procedure for description of the mechanics of infinitesimal metrically- or affinely-rigid bodies moving in non-Euclidean spaces presented in \cite{all_06} but applying it to quite new and very interesting from the geometrical point of view two-dimensional surface which is called the Mylar\footnote{According to Webster's New World Dictionary, \emph{Mylar} is a trademark for a polyester made in extremely thin sheets of great tensile strength.} balloon which is constructed by taking two circular disks of Mylar, sewing them along their boundaries and then inflating with either air or helium (see, e.g., \cite{m2004,m-o,pau}).

\begin{figure}[h]
\centering
\begin{minipage}[b]{0.45\textwidth}
\centering
\includegraphics[width=2.4in,keepaspectratio]{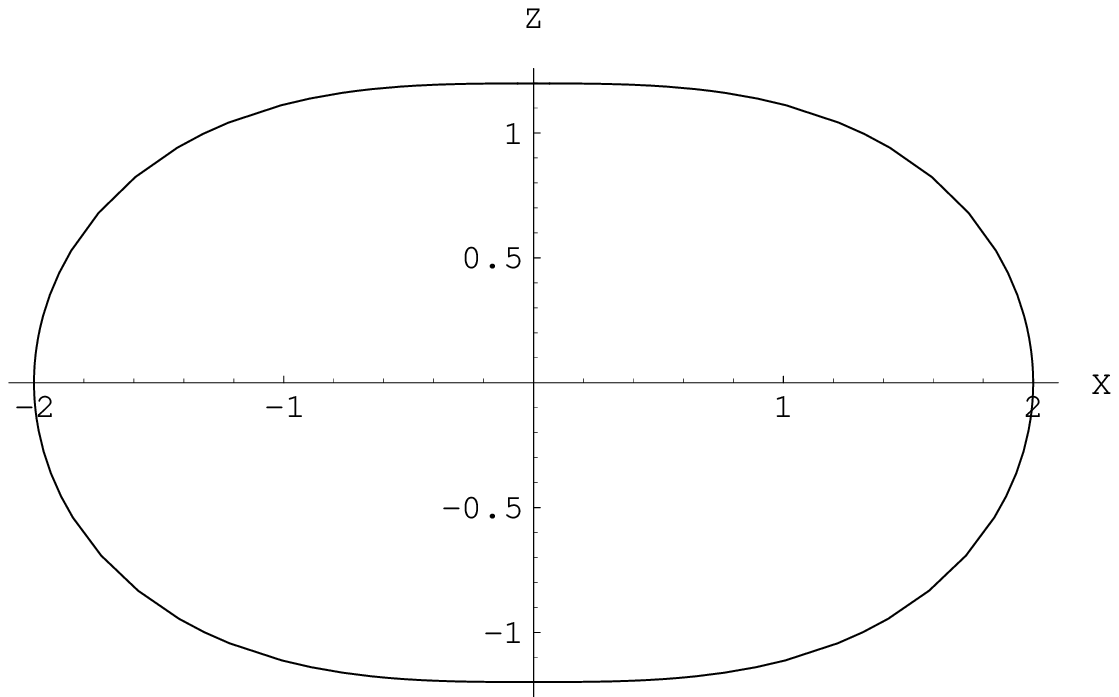}
\end{minipage}%
\hfill%
\begin{minipage}[b]{0.55\textwidth}
\vspace*{14pt} \centering
\includegraphics[width=2.8in,keepaspectratio]{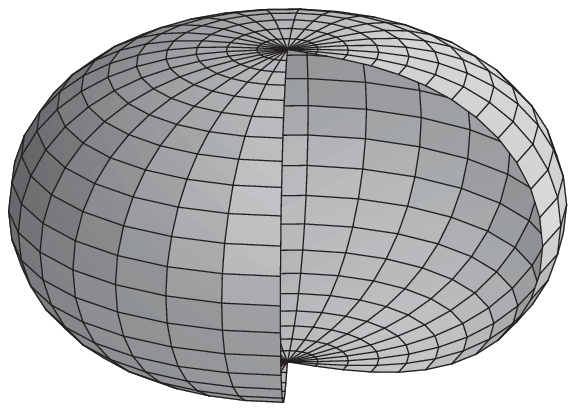}
\vspace{-0.08\textwidth}
\end{minipage}%
\end{figure}

\begin{figure}[h]
\centering
\begin{minipage}[t]{0.4\textwidth}
\caption{The profile of the Mylar balloon in $XOZ$ plane.}
\end{minipage}%
\hfill%
\begin{minipage}[t]{0.55\textwidth}
\caption{An open part of the Mylar balloon surface drawn using the
parametrization (\ref{confcoos1})--(\ref{confcoos2}).}
\end{minipage}%
\end{figure}

\section{2D Infinitesimal Gyroscope on the Mylar Balloon}\label{sec:mylar}

In conformal coordinates the Mylar balloon (see Figures 1 and 2) of the radius $r$  is
given by the following formulas:
\begin{eqnarray}
x(u,v)&=&\frac{r\cos v}{\sqrt{\cosh(2u)}},\qquad
y(u,v)=\frac{r\sin v}{\sqrt{\cosh(2u)}}\label{confcoos1} \nonumber\\
&&\\
z(u,v)&=&\sqrt{2}r[E(\arcsin (\frac{\sqrt{2}\sinh
u}{\sqrt{\cosh
(2u)}}),\frac{1}{\sqrt{2}})-\frac{1}{2}F(\arcsin(\frac{\sqrt{2}\sinh
u}{\sqrt{\cosh
(2u)}}),\frac{1}{\sqrt{2}})]\label{confcoos2}\nonumber
\end{eqnarray}
where $u\in[-\infty,\infty]$, $v\in[0,2\pi]$, and $F(z,k)$, $E(z,k)$ are the incomplete elliptic integrals of the first and second kinds, respectively.

The first and second fundamental forms are given respectively as follows (for more details see, e.g.,
\cite{m2004}):
\begin{equation}\label{eq53}
{\rm I}=\frac{r^2}{\cosh(2u)}\left(\rd u^2+ \rd v^2\right),\qquad
{\rm II}=\frac{r}{{\cosh (2u)}^{3/2}}\left(2 \rd u^2 + \rd
v^2\right).
\end{equation}
Then the metric tensor and its inverse have the following
components:
\begin{eqnarray}
&&g_{uu}=g_{vv}=\frac{r^2}{\cosh(2u)},\qquad g_{uv}=g_{vu}=0,\nonumber\label{mten1}\\
&& \\
&&g^{uu}=g^{vv}=\frac{\cosh(2u)}{r^2}, \qquad
g^{uv}=g^{vu}=0. \nonumber\label{mten2}
\end{eqnarray}
Hence, the Levi-Civita affine connection
\begin{equation}\label{eq55}
\Gamma^{i}{}_{jk}=\left\{\begin{array}{c} i \\ jk
\end{array}\right\}=
\frac{1}{2}g^{im}\left(g_{mj,k}+g_{mk,j}-g_{jk,m}\right)
\end{equation}
has only four  non-zero components
\begin{equation}\label{eq56}
\Gamma^{u}{}_{uu}=-\Gamma^{u}{}_{vv}=\Gamma^{v}{}_{uv}=\Gamma^{v}{}_{vu}=-\tanh(2u).
\end{equation}

So, for the infinitesimal gyroscope moving on the Mylar balloon, it is a natural choice of the holonomicl base to be tangent to the coordinate lines:
\begin{equation}\label{eq57}
\mathcal{E}_{u}=\frac{\partial}{\partial u},\qquad
\mathcal{E}_{v}=\frac{\partial}{\partial v},\qquad
|\mathcal{E}_{u}|=|\mathcal{E}_{v}|=\frac{r}{\sqrt{\cosh(2u)}},
\end{equation}
and then the normalized fields will form a convenient orthonormal frame
$E$ in the curved Riemannian space $(M,g)$:
\begin{equation}\label{eq58}
E_{u}=\frac{\sqrt{\cosh(2u)}}{r}\frac{\partial}{\partial u},\qquad
E_{v}=\frac{\sqrt{\cosh(2u)}}{r}\frac{\partial}{\partial
v},
\end{equation}
or written in terms of its components:
\begin{eqnarray}
E_{u}=\frac{\sqrt{\cosh(2u)}}{r}\left[\begin{array}{c}1 \\ 0\end{array}\right],
&\qquad& E_{v}=\frac{\sqrt{\cosh(2u)}}{r}\left[\begin{array}{c}0 \\ 1\end{array}\right],
\label{eq59a1} \nonumber\\
&&\\
E^{u}=\frac{r}{\sqrt{\cosh(2u)}}[1, 0],&\qquad&
E^{v}=\frac{r}{\sqrt{\cosh(2u)}}[0,1].\nonumber\label{eq59a2}
\end{eqnarray}
Then the teleparallelism connection
$\Gamma[E]^{i}{}_{jk}=E^{i}{}_{A}E^{A}{}_{j,k}$ induced by the above-described frame
$E$ has only two non-zero components
\begin{equation}\label{eq60}
\Gamma[E]^{u}{}_{uu}=\Gamma[E]^{v}{}_{vu}=-\tanh(2u).
\end{equation}
So, the affine connection $\Gamma$ in the auxiliary aholonomic
representation
\begin{equation}\label{eq61}
\Gamma^{A}{}_{BC}=E^{A}{}_{i}\left(\Gamma^{i}{}_{jk}-\Gamma[E]^{i}{}_{jk}\right)E^{j}{}_{B}E^{k}{}_{C}
\end{equation}
also has only two non-zero components
\begin{equation}\label{eq62}
\Gamma^{u}{}_{vv}=-\Gamma^{v}{}_{uv}=\frac{\sinh(2u)}{r\sqrt{\cosh(2u)}}\cdot
\end{equation}
The orthonormal frame $e=(e_{u},e_{v})$, which describes the internal configuration of our infinitesimal gyroscope moving on the Mylar balloon, is a composition of the above-introduced fixed aholonomic frame $E=(E_{u},E_{v})$ and some time-dependent orthogonal matrix $U$:
\begin{eqnarray} \label{eq63a}
e_{u}=E_{u}U^{u}{}_{u}+E_{v}U^{v}{}_{u}, \qquad
e_{v}=E_{u}U^{u}{}_{v}+E_{v}U^{v}{}_{v},
\end{eqnarray}
while $U$ can be parameterized as follows:
\begin{equation}\label{eq64}
U=\left[\begin{array}{cr}
  \cos \psi & -\sin \psi \\
  \sin \psi & \cos \psi
\end{array}\right],\qquad \widehat{\omega}_{\rm rl}= \frac{\rd \psi}{\rd t}\left[\begin{array}{cr}
  0 & -1 \\
  1 & 0
\end{array}\right]\cdot
\end{equation}
In order to calculate the ``drift'' term of the angular velocity
we should calculate at first
\begin{eqnarray}
U^{F}{}_{B}\widehat{\omega}_{\rm dr}{}^{B}{}_{A}U^{-1A}{}_{D}&=&\Gamma^{F}{}_{DC}U^{C}{}_{E}\widehat{V}^{E}\nonumber\\
&=&\Gamma^{F}{}_{DC}U^{C}{}_{E}U^{-1E}{}_{A}E^{A}{}_{i}\frac{\rd
x^{i}}{\rd t}=\Gamma^{F}{}_{DC}E^{C}{}_{i}\frac{\rd x^{i}}{\rd
t}\cdot\label{eq65}
\end{eqnarray}
Therefore, we obtain the following matrix
\begin{equation}\label{eq66}
U\widehat{\omega}_{\rm dr}U^{-1}=\tanh(2u)\frac{\rd v}{\rd
t}\left[\begin{array}{cr}
  0 & -1 \\
  1 & 0
\end{array}\right],
\end{equation}
and then
\begin{equation}\label{eq67}
\widehat{\omega}_{\rm dr}= \tanh(2u)\frac{\rd v}{\rd
t}U^{-1}\left[\begin{array}{cr}
  0 & -1 \\
  1 & 0
\end{array}\right]U= \tanh(2u)\frac{\rd v}{\rd t}\left[\begin{array}{cr}
  0 & -1 \\
  1 & 0
\end{array}\right]\cdot
\end{equation}
So, we can express the angular velocity as the sum of the ``drift'' term describing the time rate of the rotational motion contained in the field $E$ itself and ``relative'' term describing the rotational motion with respect to the fixed reference frame $E$, i.e.,
\begin{equation}\label{eq68}
\widehat{\omega}=\widehat{\omega}_{\rm dr}+\widehat{\omega}_{\rm
rl}=\left(\frac{\rd \psi}{\rd t}+\tanh(2u)\frac{\rd v}{\rd
t}\right)\left[\begin{array}{cc}
  0 & -1 \\
  1 & 0
\end{array}\right].
\end{equation}
Then the kinetic energy of our infinitesimal gyroscope moving on the Mylar balloon can be written in the following form:
\begin{eqnarray}\label{eq69}
T&=&T_{\rm
tr}+T_{\rm
int}=\frac{m}{2}g_{ij}\frac{\rd
x^{i}}{\rd
t}\frac{\rd
x^{j}}{\rd
t}+\frac{I}{2}\delta_{AB}\widehat{\omega}^{A}{}_{C}\widehat{\omega}^{B}{}_{D}\delta^{CD}\nonumber\\
&=&\frac{mr^2}{2\cosh(2u)}\left[\left(\frac{\rd
u}{\rd
t}\right)^{2}+\left(\frac{\rd
v}{\rd
t}\right)^{2}\right]
+\frac{I}{2}\left(\frac{\rd
\psi}{\rd
t}+\tanh(2u)\frac{\rd
v}{\rd
t}\right)^2,
\end{eqnarray}
where $I$ is the scalar moment of inertia of the plane rotator.

Let us now rewrite the above kinetic energy (\ref{eq69}) in the form where we have
explicitly separated the mass factor, i.e.,
\begin{equation}\label{eq70}
T=\frac{m}{2}G_{ij}(q)\frac{\rd q^{i}}{\rd t}\frac{\rd q^{j}}{\rd
t},
\end{equation}
where $\left(q^{i}\right)=(u,v,\psi)$ are the generalized coordinates and the metric matrix
$G_{ij}(q)$ is given as follows:
\begin{equation}\label{eq71}
\left[G_{ij}\right]= r^{2}\left[\begin{array}{ccccc}{\rm sech}(2u) & & 0 & & 0 \\
0 & & {\rm sech}(2u)+\displaystyle{\frac{I}{mr^{2}}}\tanh^{2}(2u) & & \displaystyle{\frac{I}{mr^{2}}}\tanh(2u) \\
0 & & \displaystyle{\frac{I}{mr^{2}}}\tanh(2u) & &
\displaystyle{\frac{I}{mr^{2}}}\end{array}\right].
\end{equation}
The square root of the determinant of the above matrix (i.e., the
weight-one volume density) is given by the following expression:
\begin{equation}\label{eq72}
\sqrt{G}=\sqrt{\det\left[G_{ij}\right]}=\sqrt{\frac{I}{m}}\
r^{2}{\rm sech}(2u).
\end{equation}
Their contravariant inverse metric $G^{ij}$ (for which
$G^{ik}G_{kj} =\delta^{i}{}_{j}$) is as follows:
\begin{equation}\label{eq73}
\left[G^{ij}\right]= \frac{1}{r^{2}}\left[\begin{array}{ccccc}\cosh(2u) & & 0 & & 0 \\
0 & & \cosh(2u) & & -\sinh(2u) \\ 0 & & -\sinh(2u) & &
\sinh(2u)\tanh(2u)+\displaystyle{\frac{m r^2}{I}}\end{array}\right].
\end{equation}

For the potential systems with Lagrangians $L=T-V(q)$ the Legendre
transformation $p_{i}=\partial L/\partial
\dot{q}^{i}=mG_{ij}(q)\dot{q}^{j}$ has the usual form:
\begin{eqnarray}
p_{u}&=&\frac{mr^{2}}{\cosh(2u)}\dot{u},\quad\quad\quad p_{\psi}=I\left[\dot{\psi}+\tanh(2u)\dot{v}\right],\label{eq74a}\\
p_{v}&=&\left[\frac{mr^{2}}{\cosh(2u)}+I\tanh^2(2u)\right]\dot{v}+I\tanh(2u)\dot{\psi}.\label{eq74c}
\end{eqnarray}
Inverting it we obtain that
\begin{eqnarray}
\dot{u}&=&\frac{\cosh(2u)}{mr^{2}}p_{u},\quad
\dot{v}=\frac{\cosh(2u)}{mr^{2}}\left(p_{v}-\tanh(2u)p_{\psi}\right),\label{eq75a}\\
\dot{\psi}&=&\frac{\sinh(2u)}{mr^{2}}\left(\left[\frac{mr^{2}}{I\sinh(2u)}+\tanh(2u)\right]p_{\psi}-p_{v}\right).\label{eq75c}
\end{eqnarray}
Now substituting (\ref{eq75a})--(\ref{eq75c}) into the expression
for the total energy,
\begin{equation}\label{eq76}
E=\dot{q}^{i}\frac{\partial L}{\partial \dot{q}^{i}}-L=T+V(q),
\end{equation}
we have that the Hamiltonian is given as
$H(q,p)=\mathcal{T}(q,p)+V(q)$. Then the geodetic Hamiltonian
(when $V(q)\equiv 0$) can be written as follows:
\begin{eqnarray}\label{eq77}
H(q,p)= \mathcal{T}(q,p)&=&\frac{\cosh(2u)}{2mr^2}\left(p_{u}^{2}+p_{v}^{2}
-2\tanh(2u)p_{v}p_{\psi}\right.\nonumber\\
&& +\left.\left[\frac{mr^2}{I\cosh(2u)}+\tanh^2(2u)\right]p_{\psi}^{2}\right).
\end{eqnarray}

\section{Hamilton-Jacobi Equation and Action-Angle Analysis}

In order to study integrability and hyperintegrability (degeneracy) problems, we need to investigate the separability of the corresponding Hamilton-Jacobi
equation expressed in the action-angle variables:
\begin{equation}\label{eq78}
\frac{\partial S}{\partial t}+H\left(q^{i},\frac{\partial
S}{\partial q^{i}}\right)=0.
\end{equation}

If we deal with time-independent problems, then the Hamilton's
principal function $S(q,t)$ can be sought in the following form:
\begin{equation}\label{eq79}
S(q,t)=S_0(q)-Et,
\end{equation}
where $E$ is the integration constant and the reduced function
$S_0$ satisfies the time-independent Hamilton-Jacobi equation:
\begin{equation}\label{eq80}
H\left(q^{i},\frac{\partial S_0}{\partial q^{i}}\right)=E.
\end{equation}

Let us note that in our case $v$ and $\psi$ are cyclic variables in the
kinetic energy term (\ref{eq77}), so we can focus our attention on
the models where the potential energy $V(q)$ also does not depend
on them, i.e., the corresponding conjugate momenta $p_v$ and
$p_{\psi}$ are constants of motion. The resulting models
(including the geodetic ones with $V(q)\equiv 0$) are completely
integrable and can be analysed with the help of the method of separation of variables.

Taking into account that we are dealing with the cyclic variables $v$ and $\psi$, the reduced
action $S_0(q)$ can be expressed in the following form:
\begin{equation}\label{eq81}
S_0(u,v,\psi;E,l,s)=S_u(u;E)+S_v(v;l)+S_\psi(\psi;s)=S_u(u;E)+lv+s\psi,
\end{equation}
where $E$, $l$, $s$ are three integration constants for the system
with three degrees of freedom (just as it should be in the complete
integral), i.e., the dependence of $S_v(v;l)$ and $S_\psi(\psi;s)$
on their arguments is postulated as linear. This means that due to
the assumed symmetry the problem is reduced to the one-dimensional
one for $S_u(u;E)$, i.e., substituting (\ref{eq81}) into
(\ref{eq80}) we obtain the following ordinary differential
equation:
\begin{eqnarray}\label{eq82}
\left(\frac{\rd S_u}{\rd u}\right)^2&=&\frac{2mr^2}{\cosh(2u)}\left(E-V(u)\right)
-l^2+2\tanh(2u)ls\nonumber\\
&& -\left(\frac{mr^2}{I\cosh(2u)}+\tanh^2(2u)\right)s^2.
\end{eqnarray}
Therefore,
\begin{eqnarray}
p_u=\frac{\partial S_0}{\partial u}=\frac{\rd S_u}{\rd u}&=&\pm\left[\frac{2mr^2}{\cosh(2u)}\left(E-V(u)\right)-l^2\right.\nonumber\\
&&-\left.\left(\frac{mr^2}{I\cosh(2u)}+\tanh^2(2u)\right)s^2+2\tanh(2u)ls\right]^{1/2}
\qquad\label{eq83a}\\
p_v=\frac{\partial S_0}{\partial v}=\frac{\rd S_v}{\rd
v}&=&l,\qquad \qquad  \qquad p_\psi=\frac{\partial S_0}{\partial
\psi}=\frac{\rd S_\psi}{\rd \psi}=s,\label{eq83b}
\end{eqnarray}
and then the corresponding action variables (i.e., the contour
integrals of the differential one-forms $p_{q^{i}}\rd q^{i}$ along
the corresponding orbits in the two-dimensional phase spaces of
the $(q^{i},p_{q^{i}})$-variables) are given as follows:
\begin{eqnarray}
J_u&=&\oint p_u \rd u=2\int^{u_{\rm max}}_{u_{\rm min}}\left[\frac{2mr^2}{\cosh(2u)}\left(E-V(u)\right)
-l^2\right.\nonumber\\
&&-\left.\left(\frac{mr^2}{I\cosh(2u)}+\tanh^2(2u)\right)s^2+2\tanh(2u)ls\right]^{1/2}\rd u,
\quad\label{eq84a}\\
J_v&=&\oint p_v\rd v=\int^{2\pi}_{0}l\rd v=2\pi l,\ J_\psi=\oint
p_\psi\rd \psi=\int^{2\pi}_{0}s\rd \psi=2\pi s,\qquad\label{eq84c}
\end{eqnarray}
where the contour integral of the differential one-form $p_{u}\rd
u$ along the corresponding orbit in the two-dimensional phase
space of the $(u,p_{u})$-variables in (\ref{eq84a}) equals twice
the integral taken between the turning points ($u_{\rm min}$,
$u_{\rm max}$ denoting the left and right turning points of the
$u$-motion, respectively) in the classically admissible region,
i.e., between the nulls of (\ref{eq82}). Substituting there
$l=J_{v}/2\pi$ and $s=J_{\psi}/2\pi$ we obtain the following
expression:
\begin{eqnarray}\label{eq85}
J_u&=&\oint\left[\frac{2mr^2}{\cosh(2u)}\left(E-V(u)\right)-\frac{J^2_{v}}{4\pi^2}\right.\nonumber\\
&&-\left.\left(\frac{mr^2}{I\cosh(2u)}+\tanh^2(2u)\right)\frac{J^2_{\psi}}{4\pi^2}
+\tanh(2u)\frac{J_{v}J_{\psi}}{2\pi^2}\right]^{1/2}\rd u.\quad
\end{eqnarray}

\subsection{Geodetic case and structure of partial degeneracy}

\noindent For the geodetic case ($V(u)\equiv 0$) we can rewrite (\ref{eq85}) as follows:
\begin{equation}\label{eq86}
J_u=\oint\sqrt{\frac{mr^2}{\cosh(2u)}\left(2E-\frac{J^2_{\psi}}{4\pi^2I}\right)-\frac{\left(J_{\psi}\tanh(2u)-J_{v}\right)^2}{4\pi^2}}\rd
u
\end{equation}
or simply
\begin{equation}\label{eq87}
J_u=\oint\sqrt{\frac{A}{\cosh(2u)}-\left(B\tanh(2u)-C\right)^2}\rd
u,
\end{equation}
where $A=mr^2\left(2E-J^2_{\psi}/4\pi^2I\right)$,
$B=J_{\psi}/2\pi$, and $C=J_{v}/2\pi$.

Next we will make  use of the well-known identities for the hyperbolic
functions
\begin{equation}\label{eq88}
{\rm sech}(2u)=\frac{1}{\cosh(2u)}=\frac{1-\tanh^2 u}{1+\tanh^2
u},\qquad \tanh(2u)=\frac{2\tanh u}{1+\tanh^2 u},
\end{equation}
and the following transformation of the independent variable:
\begin{equation}\label{eq88a}
x=\tanh u,\qquad -1\le x \le 1,\qquad \rd u = \frac{\rd x}{1-x^2}\cdot
\end{equation}
In this way we can  rewrite (\ref{eq87}) in the form appropriate
for the residue analysis:
\begin{eqnarray}\label{eq89}
J_{u}&=&\oint\sqrt{A\frac{1-x^2}{1+x^2}-\left(B\frac{2x}{1+x^2}-C\right)^2}\frac{\rd x}{1-x^2}\nonumber\\
&=&\oint\sqrt{A\left(1-x^4\right)-\left(Cx^2-2Bx+C\right)^2}\frac{\rd
x}{1-x^4}\cdot
\end{eqnarray}
According to the Cauchy's residue theorem in the complex plane we choose some positively oriented simple closed curve $\gamma$ that infinitesimally encircles the branch cut (or cuts, if there are more than one) of the complex-valued function $f(z)$. We also suppose that outside the region bounded by the curve $\gamma$ this function $f(z)$ is meromorphic, i.e., it is holomorphic on a simply connected open subset of the complex plane except for the discrete set of isolated points (including that one at infinity if it exists) $a_k$, $k=1,\ldots,n$,  which are called the poles of the function $f$. Then it can be shown that
\begin{equation}\label{eq90}
\oint_{\gamma} f(z)\rd z=-2\pi \ri \sum^{n}_{k=1}{\rm
Res}\left(f,a_{k}\right).
\end{equation}
In our case we have a complex-valued function $f$  which general
expression can be cast in the  form
\begin{equation}\label{eq92}
f(z)=-\frac{\sqrt{az^4+bz^3+cz^2+dz+e}}{(1-z)(1+z)(\ri-z)(\ri+z)}\cdot
\end{equation}
It has five poles  at $a_{k}=\{1,-1,\ri,-\ri,\infty\}$, whereas the
coefficients in (\ref{eq92}) are connected with the previously introduced ones as follows:
\begin{eqnarray}
a&=&-A-C^2=-2mr^2E+\frac{mr^2}{I}\frac{J^2_{\psi}}{4\pi^2}-\frac{J^2_{v}}{4\pi^2},\label{eq91a}\\
b&=&d=4BC=\frac{J_{\psi}J_{v}}{\pi^2}, \quad c=-4B^2-2C^2=-\frac{J^2_{\psi}}{\pi^2}-\frac{J^2_{v}}{2\pi^2}, \label{eq91b}\\
e&=&A-C^2=2mr^2E-\frac{mr^2}{I}\frac{J^2_{\psi}}{4\pi^2}-\frac{J^2_{v}}{4\pi^2}\cdot
\label{eq91c}
\end{eqnarray}
The respective residues  (\ref{eq90}) are
\begin{eqnarray}
{\rm Res}(f,1)&=&-\frac{\ri}{4}\sqrt{|a+b+c+d+e|}=-\frac{\ri}{2}|B-C|=-\frac{\ri}{4\pi}|J_{\psi}-J_{v}|,\qquad\label{eq93a}\\
{\rm Res}(f,-1)&=&-\frac{\ri}{4}\sqrt{|a-b+c-d+e|}=-\frac{\ri}{2}|B+C|=-\frac{\ri}{4\pi}|J_{\psi}+J_{v}|,\label{eq93b}\\
{\rm Res}(f,\ri)&=&\frac{\ri}{4}\sqrt{|a-\ri b-c+\ri d+e|}=\frac{\ri}{2}|B|=\frac{\ri}{4\pi}|J_{\psi}|,\label{eq93c}\\
{\rm Res}(f,-\ri)&=&\frac{\ri}{4}\sqrt{|a+\ri b-c-\ri d+e|}=\frac{\ri}{2}|B|=\frac{\ri}{4\pi}|J_{\psi}|,\label{eq93d}
\end{eqnarray}
and, because $\lim_{|z|\rightarrow\infty}f(z)=0$, the residue at
infinity is given by the formula
\begin{equation}\label{eq93e}
{\rm Res}(f,\infty)=-\lim_{|z|\rightarrow\infty}zf(z)=0.
\end{equation}
So, substituting (\ref{eq93a})--(\ref{eq93e}) into (\ref{eq90})
with the function $f(z)$ given by (\ref{eq92}) we are finally
obtaining the connection between the three action variables:
\begin{equation}\label{eq93}
2J_u=2|J_{\psi}|-|J_{\psi}-J_{v}|-|J_{\psi}+J_{v}|.
\end{equation}
Explicitly in the six regions of the phase space we have the
following structure of the partial degeneracy:
\begin{itemize}
\item[i)]
$\left(J_{\psi}>0\right)\wedge\left(J_{\psi}>J_{v}\right)\wedge\left(J_{\psi}>-J_{v}\right)$,
i.e.,
$J_{\psi}>|J_{v}|>0$,
then
\begin{equation}
J_{u}=0,
\end{equation}
\item[ii)]
$\left(J_{\psi}>0\right)\wedge\left(J_{\psi}<J_{v}\right)\wedge\left(J_{\psi}>-J_{v}\right)$,
i.e.,
$J_{v}>J_{\psi}>0$,
then
\begin{equation}
J_{u}+J_{v}-J_{\psi}=0,
\end{equation}
\item[iii)]
$\left(J_{\psi}<0\right)\wedge\left(J_{\psi}<J_{v}\right)\wedge\left(J_{\psi}>-J_{v}\right)$,
i.e.,
$-J_{v}<J_{\psi}<0$,
then
\begin{equation}
J_{u}+J_{v}+J_{\psi}=0,
\end{equation}
\item[iv)]
$\left(J_{\psi}>0\right)\wedge\left(J_{\psi}>J_{v}\right)\wedge\left(J_{\psi}<-J_{v}\right)$,
i.e.,
$-J_{v}>J_{\psi}>0$,
then
\begin{equation}
J_{u}-J_{v}-J_{\psi}=0,
\end{equation}
\item[v)]
$\left(J_{\psi}<0\right)\wedge\left(J_{\psi}>J_{v}\right)\wedge\left(J_{\psi}<-J_{v}\right)$,
i.e.,
$J_{v}<J_{\psi}<0$,
then
\begin{equation}
J_{u}-J_{v}+J_{\psi}=0,
\end{equation}
\item[vi)]
$\left(J_{\psi}<0\right)\wedge\left(J_{\psi}<J_{v}\right)\wedge\left(J_{\psi}<-J_{v}\right)$,
i.e.,
$J_{\psi}<-|J_{v}|<0$,
then
\begin{equation}
J_{u}=0.
\end{equation}
\end{itemize}
In other words, in any of the above regions only two of the three
action variables (or quantum numbers on the level of the old
quantum theory according to the Bohr-Sommerfeld postulates) are
essential.

\subsection{Non-geodetic cases with modeling potentials}

\noindent As for the non-geodetic cases, we can see that the class of integrable problems contains, for example, the following interesting potential models:
\begin{itemize}
\item[i)] the ``harmonic oscillator''-type potential model:
\begin{equation}\label{eq94}
V=\frac{\varkappa}{2}\tanh^2 u=\frac{\varkappa}{2}x^2,\qquad \varkappa>0,
\end{equation}
\item[ii)] the general ``anharmonic oscillator''-type potential model:
\begin{equation}\label{eq95}
V=\alpha x^4+\beta x^3+\gamma x^2+\delta x,\qquad \alpha>0.
\end{equation}
\end{itemize}

\subsubsection{Harmonic oscillator-type potential}

\noindent The ``harmonic oscillator''-type potential (\ref{eq94}) produces
the contour integral corresponding to (\ref{eq90}) with the
following complex-valued function:
\begin{equation}\label{eq96}
f(z)=-\frac{\sqrt{az^6+bz^5+cz^4+dz^3+ez^2+fz+g}}{(1-z)(1+z)(\ri-z)(\ri+z)},
\end{equation}
which has the same as previously poles
$a_{k}=\{1,-1,\ri,-\ri,\infty\}$ and now the coefficients in (\ref{eq96}) are connected with the previously intorduced ones as
follows:
\begin{eqnarray}
a&=&\varkappa mr^2,\quad b=0,\quad d=f=4BC=\frac{J_{\psi}J_{v}}{\pi^2},\label{eq97a}\\
c&=&-A-C^2=-2mr^2E+\frac{mr^2}{I}\frac{J^2_{\psi}}{4\pi^2}-\frac{J^2_{v}}{4\pi^2},\label{eq97b}\\
e&=&-\varkappa mr^2-4B^2-2C^2=-\varkappa mr^2-\frac{J^2_{\psi}}{\pi^2}-\frac{J^2_{v}}{2\pi^2},\label{eq97c}\\
g&=&A-C^2=2mr^2E-\frac{mr^2}{I}\frac{J^2_{\psi}}{4\pi^2}-\frac{J^2_{v}}{4\pi^2}.
\label{eq97d}
\end{eqnarray}
Calculating the values of residues in (\ref{eq96}) we are
obtaining that
\begin{eqnarray}
{\rm Res}(f,1)&=&-\frac{\ri}{4}\sqrt{|a+b+c+d+e+f+g|}\nonumber\\
&=&-\frac{\ri}{2}|B-C|=-\frac{\ri}{4\pi}|J_{\psi}-J_{v}|,\qquad\label{eq100a}\\
{\rm Res}(f,-1)&=&-\frac{\ri}{4}\sqrt{|a-b+c-d+e-f+g|}\nonumber\\
&=&-\frac{\ri}{2}|B+C|=-\frac{\ri}{4\pi}|J_{\psi}+J_{v}|,\label{eq100b}\\
{\rm Res}(f,\ri)&=&\frac{\ri}{4}\sqrt{|-a+\ri b+c-\ri d-e+\ri f+g|}\nonumber\\
&=&\frac{\ri}{2}|B|=\frac{\ri}{4\pi}|J_{\psi}|,\label{eq100c}\\
{\rm Res}(f,-\ri)&=&\frac{\ri}{4}\sqrt{|-a-\ri b+c+\ri d-e-\ri f+g|}\nonumber\\
&=&\frac{\ri}{2}|B|=\frac{\ri}{4\pi}|J_{\psi}|,\label{eq100d}
\end{eqnarray}
and again, because $\lim_{|z|\rightarrow\infty}f(z)=0$, the
residue at infinity is calculated  by using (\ref{eq93e})
\begin{equation}\label{eq100e}
{\rm
Res}(f,\infty)=-\lim_{|z|\rightarrow\infty}zf(z)=\ri\sqrt{|a|}=\ri\sqrt{\varkappa
mr^2}.
\end{equation}
So, substituting (\ref{eq100a})--(\ref{eq100e}) into (\ref{eq90})
with the function $f(z)$ given by  (\ref{eq96}) we are finally
obtaining the connection between the three action variables:
\begin{equation}\label{eq101}
2J_u=4\pi\sqrt{\varkappa mr^2}+2|J_{\psi}|-|J_{\psi}-J_{v}|-|J_{\psi}+J_{v}|.
\end{equation}
We see that again in every region only two of the three
action variables (quantum numbers in the Bohr-Sommerfeld old
quantum theory) are essential, but this time they are intertwined
with the expression $\sqrt{\varkappa mr^2}$. By the way, in the
paper \cite{mlad1} one of us have shown that the radii $r$ of the
inflated Mylar balloon can also be quantized, so we can write as well
\begin{equation}\label{eq101a}
\frac{\rm Area}{2\pi}=\frac{\pi r^2}{2}=N\in \mathbb{Z}^{+}.
\end{equation}
Therefore, in the above defined six regions of the phase space we
have the following structure of  partial degeneracy:
\begin{itemize}
\item[i)]
$\left(J_{\psi}>0\right)\wedge\left(J_{\psi}>J_{v}\right)\wedge\left(J_{\psi}>-J_{v}\right)$,
i.e.,
$J_{\psi}>|J_{v}|>0$,
then
\begin{equation}
J_{u}=2\sqrt{2\pi\varkappa mN},
\end{equation}
\item[ii)]
$\left(J_{\psi}>0\right)\wedge\left(J_{\psi}<J_{v}\right)\wedge\left(J_{\psi}>-J_{v}\right)$,
i.e.,
$J_{v}>J_{\psi}>0$,
then
\begin{equation}
J_{u}+J_{v}-J_{\psi}=2\sqrt{2\pi\varkappa mN},
\end{equation}
\item[iii)]
$\left(J_{\psi}<0\right)\wedge\left(J_{\psi}<J_{v}\right)\wedge\left(J_{\psi}>-J_{v}\right)$,
i.e.,
$-J_{v}<J_{\psi}<0$,
then
\begin{equation}
J_{u}+J_{v}+J_{\psi}=2\sqrt{2\pi\varkappa mN},
\end{equation}
\item[iv)]
$\left(J_{\psi}>0\right)\wedge\left(J_{\psi}>J_{v}\right)\wedge\left(J_{\psi}<-J_{v}\right)$,
i.e.,
$-J_{v}>J_{\psi}>0$,
then
\begin{equation}
J_{u}-J_{v}-J_{\psi}=2\sqrt{2\pi\varkappa mN},
\end{equation}
\item[v)]
$\left(J_{\psi}<0\right)\wedge\left(J_{\psi}>J_{v}\right)\wedge\left(J_{\psi}<-J_{v}\right)$,
i.e.,
$J_{v}<J_{\psi}<0$,
then
\begin{equation}
J_{u}-J_{v}+J_{\psi}=2\sqrt{2\pi\varkappa mN},
\end{equation}
\item[vi)]
$\left(J_{\psi}<0\right)\wedge\left(J_{\psi}<J_{v}\right)\wedge\left(J_{\psi}<-J_{v}\right)$,
i.e.,
$J_{\psi}<-|J_{v}|<0$,
then
\begin{equation}
J_{u}=2\sqrt{2\pi\varkappa mN}.
\end{equation}
\end{itemize}

\subsubsection{Anharmonic oscillator-type potential}

\noindent The second (more general) potential (\ref{eq95}) produces the
contour integral corresponding to (\ref{eq90}) with the following
complex-valued function:
\begin{equation}\label{eq98}
f(z)=-\frac{\sqrt{az^8+bz^7+cz^6+dz^5+ez^4+fz^3+gz^2+hz+j}}{(1-z)(1+z)(\ri-z)(\ri+z)},
\end{equation}
which has the same as previously poles
$a_{k}=\{1,-1,\ri,-\ri,\infty\}$ and now the coefficients are connected with the previously introduced ones as
follows:
\begin{eqnarray}
a&=&2\alpha mr^2,\qquad b=2\beta mr^2,\qquad c=2\gamma mr^2,\qquad d=2\delta mr^2,\label{eq99a}\\
e&=&-2\alpha mr^2-A-C^2=-2mr^2\left(\alpha+E\right)+\frac{mr^2}{I}\frac{J^2_{\psi}}{4\pi^2}-\frac{J^2_{v}}{4\pi^2},\label{eq99b}\\
f&=&-2\beta mr^2+4BC=-2\beta mr^2+\frac{J_{\psi}J_{v}}{\pi^2},\label{eq99c}\\
g&=&-2\gamma mr^2-4B^2-2C^2=-2\gamma mr^2-\frac{J^2_{\psi}}{\pi^2}-\frac{J^2_{v}}{2\pi^2}, \label{eq99d}\\
h&=&-2\delta mr^2+4BC=-2\delta mr^2+\frac{J_{\psi}J_{v}}{\pi^2},\label{eq99e}\\
j&=&A-C^2=2mr^2E-\frac{mr^2}{I}\frac{J^2_{\psi}}{4\pi^2}-\frac{J^2_{v}}{4\pi^2}\cdot
\label{eq99f}
\end{eqnarray}
Calculating the values of residues in (\ref{eq98}) we end up with
\begin{eqnarray}
{\rm Res}(f,1)&=&-\frac{\ri}{4}\sqrt{|a+b+c+d+e+f+g+h+j|}\nonumber\\
&=&-\frac{\ri}{2}|B-C|=-\frac{\ri}{4\pi}|J_{\psi}-J_{v}|,\qquad\label{eq102a}\\
{\rm Res}(f,-1)&=&-\frac{\ri}{4}\sqrt{|a-b+c-d+e-f+g-h+j|}\nonumber\\
&=&-\frac{\ri}{2}|B+C|=-\frac{\ri}{4\pi}|J_{\psi}+J_{v}|,\label{eq102b}\\
{\rm Res}(f,\ri)&=&\frac{\ri}{4}\sqrt{|a-\ri b-c+\ri d+e-\ri f-g+\ri h+j|}\nonumber\\
&=&\frac{\ri}{2}|B|=\frac{\ri}{4\pi}|J_{\psi}|,\label{eq102c}\\
{\rm Res}(f,-\ri)&=&\frac{\ri}{4}\sqrt{|a+\ri b-c-\ri d+e+\ri f-g-\ri h+j|}\nonumber\\
&=&\frac{\ri}{2}|B|=\frac{\ri}{4\pi}|J_{\psi}|,\label{eq102d}
\end{eqnarray}
and now, because this time we have that
$\lim_{|z|\rightarrow\infty}f(z)=\sqrt{a}\neq 0$, the residue at
infinity is calculated according to the formula (different from(\ref{eq93e}))
\begin{equation}\label{eq102e}
{\rm
Res}(f,\infty)=\lim_{|z|\rightarrow\infty}z^2f^{\prime}(z)=\frac{\ri}{2}\frac{|b|}{\sqrt{|a|}}=\ri |\beta|\sqrt{\frac{mr^2}{2\alpha}}.
\end{equation}
So, substituting (\ref{eq102a})--(\ref{eq102e}) into (\ref{eq90})
with the function $f(z)$ given by  (\ref{eq98}) we are finally
obtaining the connection between the three action variables:
\begin{equation}\label{eq103}
2J_u=2\pi|\beta|\sqrt{\frac{2mr^2}{\alpha}}+2|J_{\psi}|-|J_{\psi}-J_{v}|-|J_{\psi}+J_{v}|.
\end{equation}
We see again  that in every region only two of
the three action variables (quantum numbers) are essential and
they are intertwined with the quantum number $N$ corresponding to
the quantization of the radii $r$ of the inflated Mylar balloon
\cite{mlad1}. So, using (\ref{eq101a}) we can explicitly write  down the
structure of the partial degeneracy in all regions of the
phase space:
\begin{itemize}
\item[i)]
$\left(J_{\psi}>0\right)\wedge\left(J_{\psi}>J_{v}\right)\wedge\left(J_{\psi}>-J_{v}\right)$,
i.e.,
$J_{\psi}>|J_{v}|>0$,
then
\begin{equation}
J_{u}=2|\beta|\sqrt{\frac{\pi mN}{\alpha}},
\end{equation}
\item[ii)]
$\left(J_{\psi}>0\right)\wedge\left(J_{\psi}<J_{v}\right)\wedge\left(J_{\psi}>-J_{v}\right)$,
i.e.,
$J_{v}>J_{\psi}>0$,
then
\begin{equation}
J_{u}+J_{v}-J_{\psi}=2|\beta|\sqrt{\frac{\pi mN}{\alpha}},
\end{equation}
\item[iii)]
$\left(J_{\psi}<0\right)\wedge\left(J_{\psi}<J_{v}\right)\wedge\left(J_{\psi}>-J_{v}\right)$,
i.e.,
$-J_{v}<J_{\psi}<0$,
then
\begin{equation}
J_{u}+J_{v}+J_{\psi}=2|\beta|\sqrt{\frac{\pi mN}{\alpha}},
\end{equation}
\item[iv)]
$\left(J_{\psi}>0\right)\wedge\left(J_{\psi}>J_{v}\right)\wedge\left(J_{\psi}<-J_{v}\right)$,
i.e.,
$-J_{v}>J_{\psi}>0$,
then
\begin{equation}
J_{u}-J_{v}-J_{\psi}=2|\beta|\sqrt{\frac{\pi mN}{\alpha}},
\end{equation}
\item[v)]
$\left(J_{\psi}<0\right)\wedge\left(J_{\psi}>J_{v}\right)\wedge\left(J_{\psi}<-J_{v}\right)$,
i.e.,
$J_{v}<J_{\psi}<0$,
then
\begin{equation}
J_{u}-J_{v}+J_{\psi}=2|\beta|\sqrt{\frac{\pi mN}{\alpha}},
\end{equation}
\item[vi)]
$\left(J_{\psi}<0\right)\wedge\left(J_{\psi}<J_{v}\right)\wedge\left(J_{\psi}<-J_{v}\right)$,
i.e.,
$J_{\psi}<-|J_{v}|<0$,
then
\begin{equation}
J_{u}=2|\beta|\sqrt{\frac{\pi mN}{\alpha}}.
\end{equation}
\end{itemize}

\section*{Concluding Remarks}

\noindent Here we have  discussed the mechanics of the infinitesimal gyroscopes on the Mylar balloons as a two-dimensional example of
general Riemannian spaces. In  all considered cases we have found that  the corresponding Hamilton-Jacobi equation combined with the action-angle analysis leads to quite special situation when the energy $E$ does not appear in the final relationships between the action variables (contrary to the cases of other classical
surfaces, e.g., sphere, pseudo-sphere and torus \cite{jjs_76,all_06} --- and  therefore  the results concerning them deserve to be discussed separately). Nevertheless, the structure of the partial degeneracy has been obtained for the geodetic and non-geodetic situations, while for the two considered non-geodetic model potentials (the ``harmonic-oscillator"-type and general ``anharmonic-oscillator"-type ones) the action variables were also intertwined with the quantum number $N$ corresponding to the quantization of the radii $r$ of the inflated Mylar balloons. The obtained results could be applied, among others, in the theory of membranes for description  of the  motion of objects (particles) with internal structure on manifolds (e.g., transport of proteins along the curved membranes).

\section*{Acknowledgements}

\noindent The first author is very grateful to Oleksii Kostenko
from B.\ Verkin Institute for Low Temperature Physics and
Engineering of the National Academy of Sciences of Ukraine for
fruitful discussions and helpful comments on the parametrizations
of the contour integrals and corresponding calculations of the
residues of the complex-valued meromorphic functions that were
taking place during his  research visit at the Institute
of Fundamental Technological Research of the Polish Academy of
Sciences in the autumn of 2018.


\begin{thebibliography}{99}

\bibitem{all_14}
B.~Go\l ubowska, V.~Kovalchuk, J.J.~S\l awianowski, {\it
Constraints and Symmetry in Mechanics of Affine Motion}, J.
Geom. Phys. {\bf 78} (2014) 59--79.

\bibitem{all_10}
V.~Kovalchuk, {\it On Classical Dynamics of Affinely-Rigid Bodies
Subject to the Kirchhoff-Love Constraints}, Symmetry Integrability
and Geometry -- Methods and Applications {\bf 6} (2010) 031, pp
1--12.

\bibitem{m2004}
I.~Mladenov, {\it New Geometrical Applications of the Elliptic
Integrals\textup{:} The Mylar Balloon}, J. Nonlinear Math. Phys. {\bf 11},
Suppl. (2004) 55--65.

\bibitem{mlad1}
I.~Mladenov, {\it A Case Study of Quantization on Curved Surfaces\textup{:}
The Mylar Balloon}, Rendiconti del Circolo Matematico di Palermo,
Serie II, Suppl. {\bf 72} (2004) 159--169.

\bibitem{m-o}
I.~Mladenov, J.~Oprea, {\it The Mylar Balloon Revisited}, Amer.
Math. Monthly {\bf 110} (2003) 761--784.

\bibitem{pau}
W.~Paulsen, {\it What is the Shape of the Mylar Balloon?}, Amer.
Math. Monthly {\bf 101} (1994) 953--958.

\bibitem{jjs_76}
J.J.~S\l awianowski, {\it Deformable Gyroscope in a Non-Euclidean
Space. Classical Non-Relativistic Theory}, Rep. Math.
Phys. {\bf 10} (1976) 219--243.

\bibitem{all_06}
J.J.~S\l awianowski, V.~Kovalchuk, B.~Go\l ubowska, A.~Martens,
E.E.~Ro\.zko, {\it Dynamical Systems with Internal Degrees of
Freedom in Non-Euclidean Spaces}, Prace IPPT -- IFTR Reports {\bf
8}, Warsaw 2006.

\bibitem{all_17}
J.J.~S\l awianowski, V.~Kovalchuk, B.~Go\l ubowska, A.~Martens,
E.E.~Ro\.zko, {\it Mechanics of Affine Bodies. Towards Affine
Dynamical Symmetry}, J. Math. Anal. Appl. {\bf 446} (2017)
493--520.

\bibitem{all_11}
J.J.~S\l awianowski, V.~Kovalchuk, A.~Martens, B.~Go\l ubowska,
E.E.~Ro\.zko, {\it Mechanics of Systems of Affine Bodies.
Geometric Foundations and Applications in Dynamics of Structured
Media}, Mathematical Methods in the Applied Sciences {\bf 34}
(2011) 1512--1540.

\bibitem{all_12}
J.J.~S\l awianowski, V.~Kovalchuk, A.~Martens, B.~Go\l ubowska,
E.E.~Ro\.zko, {\it Essential Nonlinearity Implied by Symmetry
Group. Problems of Affine Invariance in Mechanics and Physics},
Discrete and Continuous Dynamical Systems -- Series B {\bf 17}
(2012) 699--733.

\end{thebibliography}
\end{document}